\documentclass[draft]{agujournal2019}
\usepackage{url}
\usepackage{lineno}
\usepackage[inline]{trackchanges}
\usepackage{soul}
\usepackage{amsmath}
\usepackage{amssymb}

\draftfalse

\journalname{Journal of Advances in Modeling Earth Systems}

\begin{document}

\title{Physics-Constrained Adaptive Flow Matching for Climate Downscaling}

\authors{
Kevin Debeire\affil{1,2},
Aytaç Paçal\affil{1},
Pierre Gentine\affil{3},
Luis Medrano-Navarro\affil{2},
Nils Thuerey\affil{2},
and Veronika Eyring \affil{1,4}
}
\affiliation{1}{Deutsches Zentrum für Luft- und Raumfahrt (DLR), Institut für Physik der Atmosphäre, Oberpfaffenhofen, Germany }
\affiliation{2}{School of Computation, Information and Technology, Technical University of Munich, Germany}
\affiliation{3}{LEAP Center, Columbia University, New York, NY, 10027, USA}
\affiliation{4}{Institute of Environmental Physics (IUP), University of Bremen, Bremen, Germany}
\correspondingauthor{Kevin Debeire}{kevin.debeire@dlr.de}


\begin{abstract}
Regional climate information at kilometer scales is essential for assessing the impacts of climate change, but generating it with global climate models is too expensive due to their high computational costs. Machine learning models offer a fast alternative, yet they often violate basic physical laws and degrade when applied to climates outside of their training distribution. We present Physics-Constrained Adaptive Flow Matching (PC-AFM), a generative downscaling model that addresses both problems. Building on the Adaptive Flow Matching (AFM) model of \citeA{fotiadis_stochastic_2024} as our baseline, we add soft conservation constraints that keep the downscaled output consistent with the large-scale input for precipitation and humidity, and use gradient surgery via the ConFIG algorithm to prevent these constraints from interfering with the generative objective. We train the model on Central Europe climate data, evaluate it on a $10 \times$ downscaling task (63~km to 6.3~km) over six variables (near-surface temperature, precipitation, specific humidity, surface pressure, and horizontal wind components) across a comprehensive set of metrics including bias, ensemble skill scores, power spectra, and conservation error, and test the generalization on two held-out climate regions. Within the training distribution, PC-AFM reduces conservation errors by 41\% and improves ensemble calibration while matching the baseline on standard skill metrics. Outside the training distribution, where unconstrained models develop large systematic errors by extrapolating learned statistics, PC-AFM halves precipitation wet bias, reduces conservation error by up to 52\%, and improves extreme-quantile accuracy by up to 32\%, all without any information about the target climate at inference time. Improvements extend beyond the directly constrained variables: surface pressure and temperature also benefit, suggesting that physical constraints regularize the full multivariate output rather than acting on individual variables in isolation. These results indicate that physical consistency is a practical requirement for deploying generative downscaling models in real-world applications.
\end{abstract}


\section{Introduction}
\label{sec:introduction}

While Earth system models (ESMs) have advanced considerably, their typical horizontal resolutions of 50 to 100~km remain insufficient to resolve fine-scale processes governing regional climate, particularly precipitation extremes \cite{Prein2015}. Regional climate information at kilometer-scale resolution is critical for assessing the impacts of climate change on water resources, agriculture, infrastructure, and ecosystems \cite{Giorgi2019}. Dynamical downscaling with regional climate models can reach resolutions of a few kilometers \cite{bernini_convection-permitting_nodate}, but the computational cost prohibits the large ensembles required to sample natural variability and uncertainty \cite{Rummukainen2010}. Storm-resolving global simulations such as those from the nextGEMS initiative \cite{Hohenegger2023} achieve kilometer-scale resolution globally, yet the sheer computational cost restricts them to short integration periods and small ensembles. Statistical and machine learning (ML) downscaling methods offer a computationally efficient alternative, learning a mapping from coarse to fine resolution \cite{Vandal2017, bano-medina_configuration_2020, rampal_high-resolution_2022, lin_deep_2023, rampal_enhancing_2024}.

Deterministic ML-based downscaling models optimized with Mean Squared Error (MSE) tend to produce overly smooth fields that underestimate variability and extremes, because MSE minimization favors the conditional mean \cite{Leinonen2020}. To overcome this limitation, recent work has turned to generative models, a class of ML methods that learn to sample from complex probability distributions rather than predicting a single deterministic output. Applied to downscaling, they learn the conditional distribution of high-resolution fields given coarse inputs, producing ensembles of plausible realizations that capture the stochastic nature of small-scale variability. Score-based diffusion models have emerged as a leading generative framework for atmospheric downscaling \cite{mardani_residual_2024, addison_machine_2024, hess_fast_2025, lopez-gomez_dynamical-generative_2025}. While generative models produce realistic and variable fields, they offer no guarantee that the generated samples satisfy fundamental physical properties such as mass or energy conservation. Incorporating such physical constraints into generative models is, however, non-trivial. Several approaches have been proposed to enforce such constraints within diffusion models \cite{shu_physics-informed_2023,amoros-trepat_guiding_2026}. Flow matching \cite{Lipman2023} and stochastic interpolants \cite{Albergo2023} offer a cleaner foundation for incorporating physical constraints. Rather than reversing a diffusion process, flow matching trains a neural network to transport samples from a simple base distribution to the target data distribution along smooth, direct trajectories. The denoiser is trained to predict the clean output directly at each step, providing a natural point at which conservation constraints can be evaluated on the predicted output \cite{baldan_physics_2026} \citeA{fotiadis_stochastic_2024} extended this framework for downscaling with CorrDiff\texttt{++}, by adding a learned encoder that maps the coarse input to an initial high-resolution estimate. Their Adaptive Flow Matching (AFM) model matches or exceeds diffusion-based downscaling skill while requiring fewer inference steps.

Geographic generalization has emerged as a critical open challenge for generative downscaling: models trained on a single region degrade when applied to climatologically distinct locations, as demonstrated by systematic benchmarks across precipitation regimes
\cite{harder_rainshift_2025}. Global transferability studies have shown that no current generative model maintains skill across all geographic contexts \cite{harder_global_2025}. A key mechanism underlying this failure is the violation of physical conservation laws: generative models trained on one climate regime implicitly encode the area-averaged statistics of that regime into their learned representations, and these tendencies break down under
distributional shift. \citeA{saccardi_assessing_2025} document this empirically, showing that models trained on Central Europe fail to maintain physical consistency when transferred to climatologically distinct regions such as Iberia and Scandinavia. \citeA{baldan_physics_2026} established the existence of a Pareto-optimal trade-off between distributional accuracy and constraint satisfaction in flow matching: improving physical consistency comes at the cost of some degradation in distributional fidelity, and vice versa. This result implies that physics constraints cannot simply be added for free, and that their integration requires a principled multi-objective optimization strategy. In practice, gradients from the reconstruction and physics loss terms can conflict, leading to suboptimal outcomes if summed. The ConFIG (Conflict-Free Gradient) algorithm \cite{Liu2024} offers a principled way to navigate this trade-off by computing a combined update that makes non-negative progress on all objectives simultaneously, steering the optimization toward the Pareto front without manual loss weight tuning.

In this work, we introduce Physics-Constrained Adpative Flow Matching (PC-AFM), which extends the AFM architecture of \citeA{fotiadis_stochastic_2024} with soft conservation losses and ConFIG gradient surgery to produce physically consistent downscaled fields. Our specific contributions are: (i) the introduction of physics-constrained conservation losses penalizing deviations between the coarsened prediction and the coarse-scale input for precipitation and specific humidity, evaluated in physical space with cosine-latitude weighting; (ii) a ConFIG gradient surgery to resolve conflicts between the generative and conservation objectives; (iii) a noise-level-dependent loss weighting and warmup schedule to stabilize multi-objective training; and (iv) comprehensive evaluation on the nextGEMS cycle~3 simulations, downscaling six variables by $10\times$ with an ESMValTool-based diagnostic suite (Earth System Model Evaluation Tool, \citeA{Righi2020,eyring_taking_2019}) over the training domain (Central Europe) and two withheld climate regions (Iberia and Northern Europe).

The remainder of this paper is organized as follows. Section~\ref{sec:data} describes the data and preprocessing. Section~\ref{sec:methods} presents the model. Section~\ref{sec:evaluation} details the evaluation protocol. Section~\ref{sec:results} reports results. Section~\ref{sec:summary} provides a summary of this study.


\section{Data}
\label{sec:data}

\subsection{nextGEMS Cycle 3 Simulations}
\label{sec:data_nextgems}

We use cycle~3 simulations from the ICON model within the nextGEMS project \cite{Hohenegger2023,stevens_dyamond_2019}, which provides 5.5 years of global climate simulation at approximately 6.3~km horizontal resolution. These convection-permitting simulations explicitly resolve deep convective processes rather than relying on parameterization, yielding more realistic precipitation distributions, particularly for extremes as documented in \citeA{wille_extreme_2025}. Low-resolution inputs are derived from the same high-resolution fields by cosine-latitude-weighted spatial averaging (Equation~\ref{eq:coarsening}), ensuring physical consistency between the coarse and fine scales by construction.

\subsection{Variables and Preprocessing}
\label{sec:data_variables}
The downscaling task maps 37 input channels to six high-resolution surface output variables (Table~\ref{tab:variables}). Each training sample is a $320 \times 320$ grid-point patch at the native 6.3~km resolution. Let $x_{m,n}^{\text{HR}}$
denote the value of a field at high-resolution grid cell $(m, n)$, where $m$ and $n$
index latitude and longitude respectively. Low-resolution inputs are constructed by
coarsening with a factor of $f = 10$, yielding $32 \times 32$ patches at 63~km,
using cosine-latitude-weighted spatial averaging:

\begin{linenomath*}
\begin{equation}
\label{eq:coarsening}
    \bar{x}_{i,j}^{\text{LR}} = \frac{\displaystyle\sum_{m,n \in \mathcal{P}_{ij}}
    x_{m,n}^{\text{HR}} \cos(\phi_{m})}{\displaystyle\sum_{m,n \in \mathcal{P}_{ij}}
    \cos(\phi_{m})}
\end{equation}
\end{linenomath*}
where $(i,j)$ indexes the coarse-resolution grid cells, $\mathcal{P}_{ij}$ is the set
of high-resolution cells $(m,n)$ contained within coarse cell $(i,j)$, and $\phi_m$ is
the latitude of row $m$. The low-resolution input is bilinearly upsampled to the high-resolution grid and concatenated with the static invariant fields to form the 37-channel input tensor. Input and output variables are normalized to zero mean and unit variance using training-set statistics.

\begin{table}[ht]
\caption{Input and output variables. The input comprises 37 channels: 10 surface fields, 24 pressure-level fields, and 3 static invariants. The output comprises 6 surface variables.}
\label{tab:variables}
\centering
\begin{tabular}{lll}
\hline
\textbf{Category} & \textbf{Variables} & \textbf{Levels} \\
\hline
\multicolumn{3}{l}{\textit{Surface input fields (10 variables)}} \\
 & \texttt{tas}, \texttt{uas}, \texttt{vas}, \texttt{pr} & Surface \\
 & \texttt{huss}, \texttt{ps} & Surface \\
 & \texttt{rlds}, \texttt{rlus}, \texttt{rsds}, \texttt{rsus} & Surface \\
\hline
\multicolumn{3}{l}{\textit{Pressure-level input fields (24 variables)}} \\
 & \texttt{ta}, \texttt{ua}, \texttt{va}, \texttt{zg} & 850, 700, 500, 300, 200~hPa \\
 & \texttt{hus} & 850, 700, 500~hPa \\
 & \texttt{air\_pressure} & 850~hPa \\
\hline
\multicolumn{3}{l}{\textit{Static invariant fields (3 variables)}} \\
 & \texttt{elev\_mean} (orography), \texttt{lsm\_mean} (land-sea mask),\\ & \texttt{latitude} & -- \\
\hline
\multicolumn{3}{l}{\textit{Output fields (6 variables)}} \\
 & \texttt{uas}, \texttt{vas} & 10~m wind components [m~s$^{-1}$] \\
 & \texttt{tas} & 2~m temperature [K] \\
 & \texttt{pr} & Precipitation flux [kg~m$^{-2}$~s$^{-1}$] \\
 & \texttt{huss} & Specific humidity, lowest model level [kg~kg$^{-1}$] \\
 & \texttt{ps} & Surface pressure [Pa] \\
\hline
\end{tabular}
\end{table}

\subsection{Training and Evaluation Domains}
\label{sec:data_splits}

\begin{figure}[htbp]
    \centering
    \includegraphics[width=0.8\textwidth]{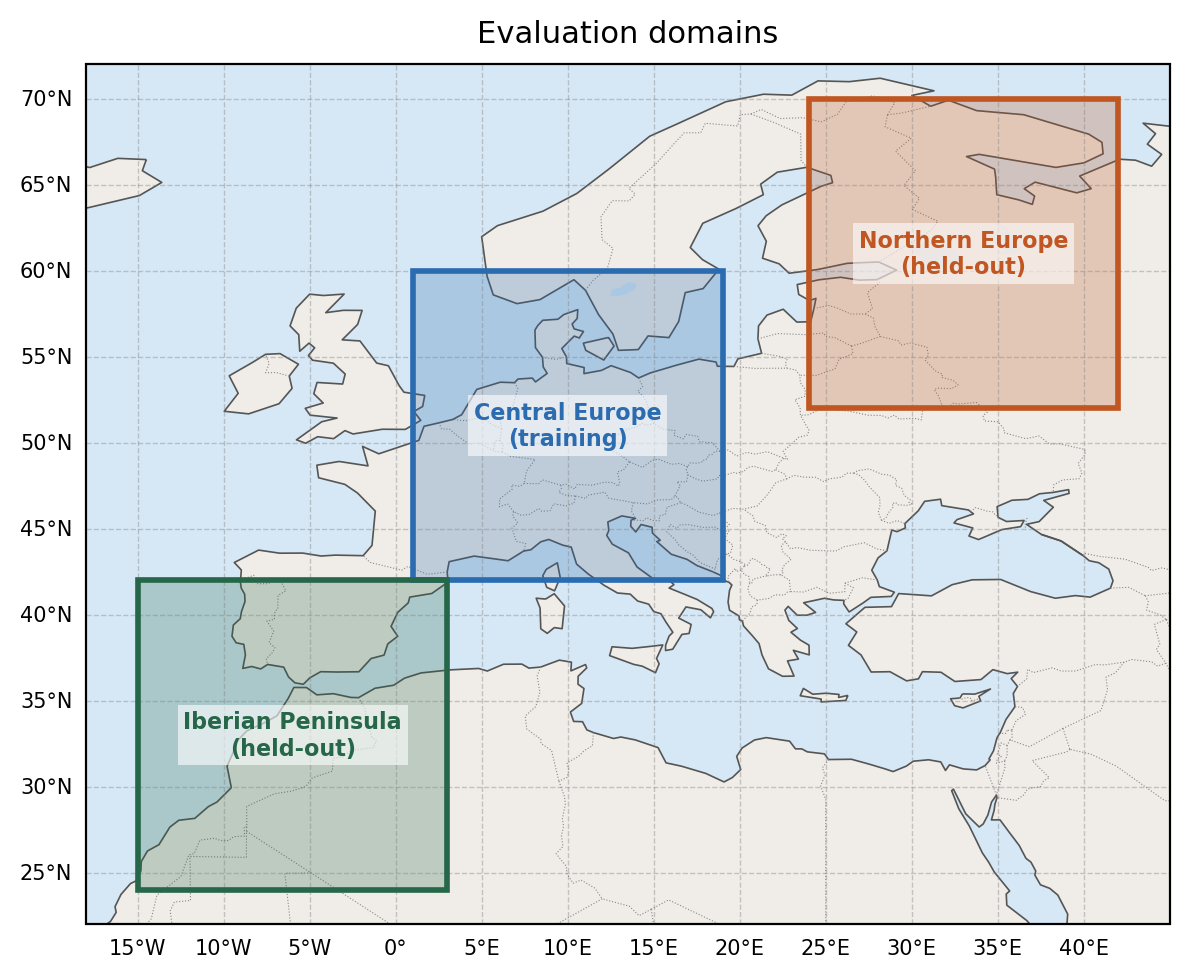}
    \caption{Geographic domains used for training and evaluation. Evaluation diagnostics for all three domains are computed using ESMValTool. The Central Europe domain (blue) is used for training. The Iberian Peninsula (green) and Northern Europe region (orange) are withheld from training and used exclusively for out-of-distribution evaluation.}
    \label{fig:domains}
\end{figure}

Data are split in a temporally contiguous way into training, validation, and test periods. The training period covers 20~January~2020 to 19~January~2024 at 3-hourly resolution, providing 11680 time steps. The test period spans the full year immediately following training
(20~January~2024 to 20~January~2025, providing 2920 time steps), chosen to cover
one complete annual cycle so that all seasons are equally represented in the evaluation.
The validation period uses the subsequent available simulation data
(21~January~2025 to 20~July~2025) used for checkpoint selection during
training.
All models are trained on the Central Europe domain (1--19°E, 42--60°N). To assess generalization beyond the training distribution, we evaluate this generalization on two geographically and climatologically distinct regions that do not overlap with the Central Europe training domain: (i) the Iberian Peninsula and Northern Morocco (15°W--3°E, 24--42°N, semi-arid Mediterranean to arid) and (ii) Northern Europe (24--42°E, 52--70°N humid subarctic to temperate maritime). The three non-overlapping domains are shown in Figure~\ref{fig:domains}. The evaluation tests geographic but not temporal out-of-distribution generalization. Temporal generalization under transient climate change is discussed in Section~\ref{sec:summary}.

\section{Methods}
\label{sec:methods}

\begin{figure}[htbp]
    \centering
    \includegraphics[width=\textwidth]{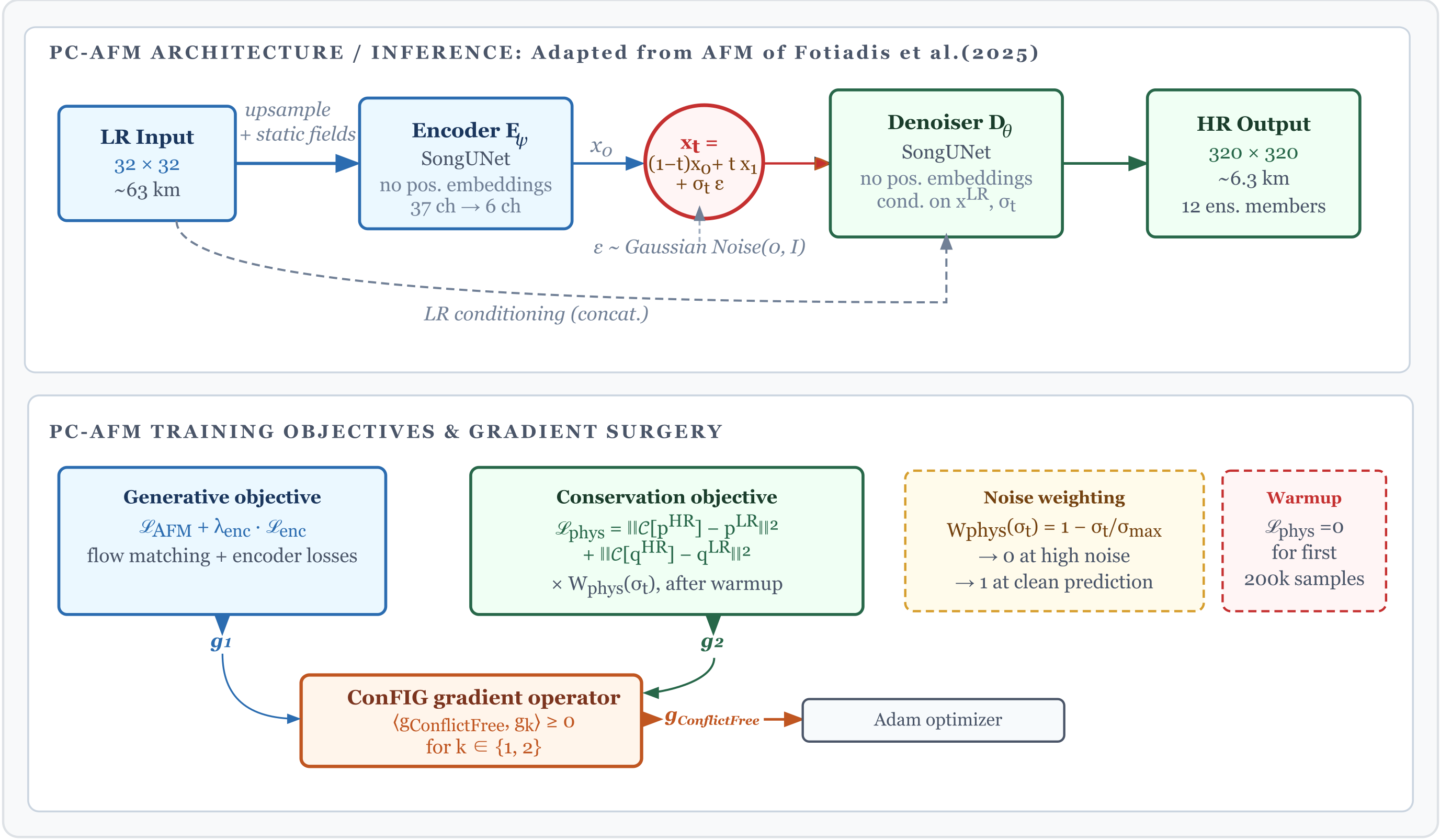}
    \caption{Overview of the PC-AFM architecture and training procedure. \textbf{Top:} At inference, the low-resolution input (32$\times$32, 63~km) is bilinearly upsampled and passed through the learned encoder $E_\psi$ to produce an initial high-resolution estimate $\hat{x}_0$. A stochastic interpolant $x_t = (1-t)\hat{x}_0 + t x_1 + \sigma_t \varepsilon$ is constructed and refined by the denoiser $D_\theta$, conditioned on the low-resolution input and noise level $\sigma_t$. Fifty denoising steps yield 12 ensemble members at 320$\times$320 resolution (6.3~km). Both networks use no positional embeddings to support geographic generalization. \textbf{Bottom:} During training, gradients from the generative objective ($\mathcal{L}_\text{AFM} + \lambda_\text{enc}\,\mathcal{L}_\text{enc}$) and the physics-constrained conservation objective ($\mathcal{L}_\text{phys}$) are combined via the ConFIG operator, which ensures non-negative progress on both objectives simultaneously. The conservation penalty is down-weighted at high noise levels and disabled during a 200k-sample warmup period.}
    \label{fig:architecture}
\end{figure}

PC-AFM builds directly on the adaptive flow matching (AFM) architecture of \citeA{fotiadis_stochastic_2024}, also known as CorrDiff\texttt{++}. The AFM model consists of two components: a deterministic encoder $E_\psi$ that maps the coarse input to a structured high-resolution initialization, and a stochastic denoiser $D_\theta$ that refines this initialization along a learned probability path to produce high-resolution ensemble members. We adopt the encoder-denoiser architecture and generative training objective of \citeA{fotiadis_stochastic_2024} without modification, with one exception: we disable positional embeddings in both networks to enable geographic generalization (Section~\ref{sec:AFM}). PC-AFM adds three training-time components to this backbone: (i) soft conservation losses that penalize deviations between the coarsened prediction and the coarse-scale input for precipitation and specific humidity
(Section~\ref{sec:physics_losses}); (ii) ConFIG gradient surgery to prevent these losses from conflicting with the generative objective (Section~\ref{sec:config}); and (iii) a noise-level-dependent weighting and warmup schedule to stabilize multi-objective training (Section~\ref{sec:physics_losses}). An overview is shown in Figure~\ref{fig:architecture}.

\subsection{Adaptive Flow Matching Backbone}
\label{sec:AFM}
Flow matching \cite{Lipman2023} trains a neural network to transport samples from a
base distribution $p_0$ to the data distribution $p_1$ along smooth probability paths
parameterized by a time-varying velocity field. Following \citeA{fotiadis_stochastic_2024}, we replace the isotropic Gaussian base distribution with the output of a learned encoder $E_\psi$ that maps the upsampled low-resolution input to an initial high-resolution estimate. This structured initialization captures large-scale spatial structure so that the generative refinement focuses on realistic fine-scale detail. The stochastic interpolant between the encoder output and the target $x_1$ is:
\begin{linenomath*}
\begin{equation}
\label{eq:interpolant}
    x_t = (1-t)\, E_\psi(x^{\text{LR}}) + t\, x_1 + \sigma_t\, \varepsilon, \quad
    \varepsilon \sim \mathcal{N}(0, I), \quad \sigma_t \sim \mathcal{U}[0, \sigma_{\max}],
\end{equation}
\end{linenomath*}
$\sigma_{\max}$ is a per-channel learnable parameter initialized at 1.0 and smoothed via exponential moving average during training. $t = 1 - \sigma_t / \sigma_{\max}$ links the continuous time variable $t \in [0,1]$ to the noise level $\sigma_t \in [0, \sigma_{\max}]$. Such that at $t=0$ (i.e.\ $\sigma_t = \sigma_{\max}$), the interpolant is a noisy version of the encoder output, and at $t=1$ (i.e.\ $\sigma_t = 0$), it recovers the clean target $x_1$. Following \citeA{karras_elucidating_2022}, the denoiser $D_\theta$ predicts the clean target $x_1$ directly from the noisy interpolant. The AFM training loss is:
\begin{linenomath*}
\begin{equation}
\label{eq:AFM_loss}
    \mathcal{L}_{\text{AFM}} = \mathbb{E}_{t, \varepsilon} \left[ w(\sigma_t)\, \left\| D_\theta(x_t, \sigma_t, x^{\text{LR}}) - x_1 \right\|^2 \right], \quad w(\sigma_t) = \frac{\sigma_t^2 + \sigma_{\text{data}}^2}{(\sigma_t \cdot \sigma_{\text{data}})^2}
\end{equation}
\end{linenomath*}
with $\sigma_{\text{data}} = 0.5$. An auxiliary encoder loss $\mathcal{L}_{\text{enc}} = \| E_\psi(x^{\text{LR}}) - x_1 \|^2$ encourages the encoder to produce a good deterministic first estimate. The baseline objective is $\mathcal{L}_{\text{base}} = \mathcal{L}_{\text{AFM}} + \lambda_{\text{enc}}\, \mathcal{L}_{\text{enc}}$. Following \citeA{fotiadis_stochastic_2024}, we adopt $\lambda_{\text{enc}} = 0.25$, which was found to yield the best performance in their ablation analysis.

At inference, $\sigma_t$ is discretized over $N$ steps following the 
\citeA{karras_elucidating_2022} schedule:
\begin{linenomath*}
\begin{equation}
\label{eq:sigma_schedule}
    \sigma_n = \left( \sigma_{\max}^{1/\rho} + \frac{n}{N-1}
    \left( \sigma_{\min}^{1/\rho} - \sigma_{\max}^{1/\rho} \right) \right)^{\rho},
    \quad \rho = 7, \quad \sigma_{\min} = 0.002.
\end{equation}
\end{linenomath*}
A trajectory is initialized as $x_{t_0} = E_\psi(x^{\text{LR}}) + \sigma_{\max}\,\varepsilon$, $\varepsilon \sim \mathcal{N}(0, I)$. At each step $n \in \{0, \ldots, N-1\}$, the denoiser predicts the clean field $\hat{x}_1 = D_\theta(x_{t_n}, \sigma_n, x^{\text{LR}})$ and the velocity is estimated as $u_{t_n} = (\hat{x}_1 - x_{t_n}) / t_n$. The state is then advanced by a single Euler step $x_{t_{n+1}} = x_{t_n} + u_{t_n}\,\Delta t$, where $\Delta t = t_n - t_{n+1} > 0$. Although $D_\theta$ predicts $x_1$ directly at every step, the Euler integration is necessary as each prediction is only locally accurate. The $N$ steps progressively refine the trajectory from the noisy initialization to the clean output. We choose $N=50$ steps similar to \citeA{fotiadis_stochastic_2024}. Repeating the process from $N_{ens}$ independent noise draws of $\varepsilon$ produces $N_{ens}$ ensemble members.

The encoder $E_\psi$ is a SongUNet \cite{Song2021} mapping the 37-channel input to the 6-channel output space with channel multipliers [1, 2, 2, 4, 4]. This configuration is referred to as SongUnet-L in \citeA{fotiadis_stochastic_2024}. The denoiser $D_\theta$ is a similar SongUNet \cite{Song2021} conditioned on the noisy interpolant $x_t$, noise level $\sigma_t$ (via Fourier embedding), and the upsampled low-resolution input $x^{\text{LR}}$. One architectural difference from \citeA{fotiadis_stochastic_2024} is that we disable positional embeddings in both the encoder and denoiser. Positional embeddings learned on a fixed training domain encode location-specific spatial structure that does not transfer to held-out regions. By removing them, the model is forced to rely solely on the physical input fields and static invariants (orography, land-sea mask, latitude) to infer spatial context.

\subsection{Physics-Constrained Conservation Losses}
\label{sec:physics_losses}

For extensive quantities such as precipitation and water vapor, the area-weighted spatial average of the downscaled field over any coarse grid cell should equal the corresponding coarse-scale input:

\begin{linenomath*}
\begin{equation}
\label{eq:conservation}
    \mathcal{C}[\hat{x}^{\text{HR}}]_{i,j} = x^{\text{LR}}_{i,j}, \quad \forall\, (i,j)
\end{equation}
\end{linenomath*}
where $\hat{x}^{\text{HR}} = D_\theta(x_t, \sigma_t, x^{\text{LR}})$ is the predicted high-resolution field, and $\mathcal{C}[\cdot]$ is the cosine-latitude-weighted coarsening operator of Equation~\ref{eq:coarsening}. We enforce this as a soft penalty. After denormalizing the denoiser output to physical units, the combined conservation loss over precipitation and specific humidity is:

\begin{linenomath*}
\begin{equation}
\label{eq:cons_loss}
    \mathcal{L}_{\text{phys}} = \left\| \mathcal{C}[\hat{p}^{\text{HR}}] - p^{\text{LR}} \right\|^2 + \left\| \mathcal{C}[\hat{q}^{\text{HR}}] - q^{\text{LR}} \right\|^2.
\end{equation}
\end{linenomath*}
where $\hat{p}^{\text{HR}}$ and $\hat{q}^{\text{HR}}$ denote the denormalized predicted precipitation and specific humidity fields, and $p^{\text{LR}}$, $q^{\text{LR}}$ are the corresponding coarse-scale inputs.
Evaluating the constraint in physical (denormalized) space ensures the penalty is proportional to physically meaningful deviations and invariant to normalization choices, making it well-suited for transfer to new climate regions. To focus the penalty on near-clean predictions where it is semantically meaningful, we apply a noise-level-dependent down-weighting:

\begin{linenomath*}
\begin{equation}
\label{eq:phys_weight}
    w_{\text{phys}}(\sigma_t) = 1 - \frac{\sigma_t}{\sigma_{\max}}
\end{equation}
\end{linenomath*}
which ramps from zero at maximum noise to one at zero noise. We additionally disable $\mathcal{L}_{\text{phys}}$ during the first $N_{\text{warmup}} = 200{,}000$ training samples, allowing the model to establish a reasonable mapping before the conservation constraints are involved in the training.

\subsection{Multi-Objective Optimization with Gradient Surgery}
\label{sec:config}

Adding $\mathcal{L}_{\text{phys}}$ creates a multi-objective problem in which conservation gradients can conflict with those from $\mathcal{L}_{\text{base}}$. Naive weighted summation provides no mechanism to detect or resolve such conflicts: at low $\lambda_{\text{phys}}$ the conservation objective is suppressed; at high $\lambda_{\text{phys}}$ the generative quality degrades. We employ ConFIG \cite{Liu2024}, which eliminates this hyperparameter by operating directly on the gradient directions rather than the loss magnitudes. At each step, ConFIG computes separate gradients $g_1$ (from $\mathcal{L}_{\text{base}}$) and $g_2$ (from $\mathcal{L}_{\text{phys}}$) and produces a conflict-free gradient update $g_{\text{CF}}$ satisfying:

\begin{linenomath*}
\begin{equation}
\label{eq:config_condition}
    \langle g_{\text{CF}},\, g_k \rangle \geq 0, \quad k \in \{1, 2\}.
\end{equation}
\end{linenomath*}

This guarantees non-negative progress on both objectives simultaneously without manual tuning of loss weights.

\subsection{Training and Inference Configuration}
\label{sec:training}

All models are trained with Adam \cite{Kingma2015} (learning rate $2 \times 10^{-4}$, $\beta_1 = 0.9$, $\beta_2 = 0.999$, batch size 256) for $6$ million samples. An exponential moving average of the denoiser parameters is maintained with a half-life of $5 \times 10^5$ images, following \citeA{karras_elucidating_2022}. At inference time, we generate $N_{ens}=12$ independent ensemble members per input time step.


\section{Evaluation Protocol}
\label{sec:evaluation}

We implement a comprehensive evaluation suite as an ESMValTool \cite{Righi2020,eyring_taking_2019} diagnostic. All metrics are computed on the test period (20~January~2024 to 20~January~2025) after conversion to standard physical units: temperature in $^\circ$C, precipitation in mm~h$^{-1}$, specific humidity in g~kg$^{-1}$, surface pressure in hPa, and wind components in m~s$^{-1}$. The one-year test period, while short, spans a full annual cycle and is consistent with the available nextGEMS cycle~3 simulation length.

\subsection{Baselines}
\label{sec:baselines}

To focus on the incremental contribution of physics constraints over the same generative backbone, we compare PC-AFM against AFM \cite{fotiadis_stochastic_2024}, the unconstrained flow matching baseline ($\mathcal{L}_{\text{phys}} = 0$) with the same architecture as PC-AFM. Unless stated otherwise, all relative performance ratios in the results section are PC-AFM / AFM.

\subsection{Evaluation Metrics}
\label{sec:metrics}

We evaluate using the following diagnostics, drawing on the metric selection of \citeA{lopez-gomez_dynamical-generative_2025} and \citeA{schillinger_enscale_2025}.

\textbf{Bias and Continuous Ranked Probability Score (CRPS):} the time-mean pointwise bias is
\begin{linenomath*}
\begin{equation}
\label{eq:bias}
    \text{Bias}_{i,j} = \frac{1}{T} \sum_{t=1}^{T} \left( \bar{\hat{x}}_{t,i,j}^{\text{HR}} - x_{t,i,j}^{\text{HR}} \right),
\end{equation}
\end{linenomath*}
where $\bar{\hat{x}}_{t}^{\text{HR}}$ denotes the ensemble-mean prediction at time step $t$. Persistent positive or negative bias indicates that the model systematically over- or underestimates the climatological mean of a variable. For precipitation, we additionally report the relative bias, normalized by the local time-mean reference value. The continuous ranked probability score \cite{Gneiting2007} evaluates the full predictive distribution against the verification value and jointly rewards calibration and sharpness:
\begin{linenomath*}
\begin{equation}
\label{eq:crps}
    \text{CRPS}(F, x) = \int_{-\infty}^{\infty} \left( F(y) - \mathbf{1}[y \geq x] \right)^2 \mathrm{d}y,
\end{equation}
\end{linenomath*}
where $F$ is the predictive cumulative distribution function estimated from the 12 ensemble members and $x$ is the reference value. A lower CRPS indicates that the ensemble is both accurate and appropriately spread around the truth.

\textbf{Relative Average Log Spectral Distance (RALSD):} the RALSD quantifies spectral fidelity by comparing radially averaged power spectral densities $S_{\text{ref}}(k)$ and $S_{\text{pred}}(k)$ at discrete wavenumber $k$ \cite{schillinger_enscale_2025}, where $S_{\text{pred}}(k)$ is computed from the spectrum of each ensemble member and averaged across members before entering the formula:
\begin{linenomath*}
\begin{equation}
\label{eq:ralsd}
    \text{RALSD} = \sqrt{\frac{1}{K}\sum_{k=1}^{K} \left(10 \log_{10} \frac{S_{\text{ref}}(k)}{S_{\text{pred}}(k)}\right)^2}.
\end{equation}
\end{linenomath*}
RALSD captures whether the model adds realistic spatial structure at the downscaled resolution, independently of pointwise accuracy.

\textbf{Conservation error (CE):} the CE measures cross-scale consistency as the time-mean absolute difference between the coarsened ensemble-mean prediction and the coarsened reference at each coarse grid cell $(i,j)$, using the operator $\mathcal{C}[\cdot]$ defined in Equation~\ref{eq:coarsening}:
\begin{linenomath*}
\begin{equation}
\label{eq:conservation_error}
    \text{CE}_{i,j} = \frac{1}{T}\sum_{t=1}^{T} \left| \mathcal{C}[\bar{\hat{x}}_t^{\text{HR}}]_{i,j} - \mathcal{C}[x_t^{\text{HR}}]_{i,j} \right|.
\end{equation}
\end{linenomath*}
A large conservation error indicates that the fine-scale prediction, when spatially aggregated, drifts away from the large-scale input it was conditioned on, a signal that the model has failed to maintain cross-scale physical consistency. The conservation error can only be calculated for variables with explicit conservation constraint (\texttt{pr} and \texttt{huss}).

\textbf{Log-PDF distance:} let $p_{\text{ref}}$ and $p_{\text{pred}}$ denote the normalized marginal histograms of the reference and ensemble-mean prediction, pooled over all grid points and time steps in the test period. The log-probability density function (log-PDF) distance is the root-mean-square deviation
between their log$_{10}$-transformed bin values. Because the comparison is performed in log space, errors in the low-probability tails contribute more strongly than errors near the mode, making this metric sensitive to misrepresentation of rare but impactful events.

\textbf{Ensemble Miscalibration (MCB):} the MCB is derived from the rank histogram (globally pooled): for each reference value, we record the rank of $x_t^{\text{HR}}$ within the sorted ensemble of $M = 12$ members. A perfectly calibrated ensemble produces a uniform rank histogram. MCB quantifies the deviation from uniformity \cite{schillinger_enscale_2025}:
\begin{linenomath*}
\begin{equation}
\label{eq:mcb}
    \text{MCB} = \frac{1}{M+1} \sum_{r=1}^{M+1} \left( h_r - \frac{1}{M+1} \right)^2,
\end{equation}
\end{linenomath*}
where $h_r$ is the observed relative frequency of rank $r$. U-shaped rank histograms with high MCB indicate an underdispersed ensemble that is overconfident. Dome-shaped histograms indicate overdispersion.

\textbf{Quantile Mean Absolute Error (MAE):} following \citeA{lopez-gomez_dynamical-generative_2025}, we assess distributional accuracy for impact-relevant compound variables through seasonal quantile diagnostics. At each grid point, we compute the empirical quantile functions $Q_{\text{ref}}(\tau)$ and $Q_{\text{pred}}(\tau)$ of the reference and ensemble-mean prediction over a seasonal subset of the test period. The quantile MAE is then averaged spatially across the domain:
\begin{linenomath*}
\begin{equation}
\label{eq:quantile_mae}
    \text{Quantile MAE} = \frac{1}{|\Omega|} \sum_{i,j \in \Omega} \frac{1}{|\mathcal{T}|} \sum_{\tau \in \mathcal{T}} \left| Q_{\text{ref}}^{i,j}(\tau) - Q_{\text{pred}}^{i,j}(\tau) \right|,
\end{equation}
\end{linenomath*}
where $\Omega$ denotes the set of grid points and $\mathcal{T}$ the set of quantile levels. Unlike pointwise metrics, quantile MAE measures how accurately the model reproduces the shape of the marginal distribution across its full range, and is therefore sensitive to errors in both the bulk and the tails. We apply this diagnostic to three seasonal composites: summer (JJA) wet-bulb globe temperature (WBGT, derived from \texttt{tas}, \texttt{huss}, and \texttt{ps} following the simplified Liljegren formula as a heat stress indicator), winter (DJF) precipitation, and fall (SON) wind speed ($v_{10} = \sqrt{\texttt{uas}^2 + \texttt{vas}^2}$). For WBGT and wind speed, quantiles span the full distribution, $\mathcal{T} = \{0.01, 0.02, \ldots, 0.99\}$; for precipitation, we restrict to the upper tail, $\mathcal{T} = \{0.95, 0.96, \ldots, 0.99\}$, to focus on extremes relevant to flood hazard. These composites target the seasons and variables most relevant to heat, flood, and wind hazard assessment respectively.


\section{Results}
\label{sec:results}


We evaluate PC-AFM against the AFM baseline across three geographic domains: the training domain (Central Europe) and the two held-out regions (Iberia/Morocco domain and Northern Europe).

\subsection{Overall Performance on Central Europe}
\label{sec:results_overall}

\begin{figure}
    \centering
    \includegraphics[width=0.85\textwidth]{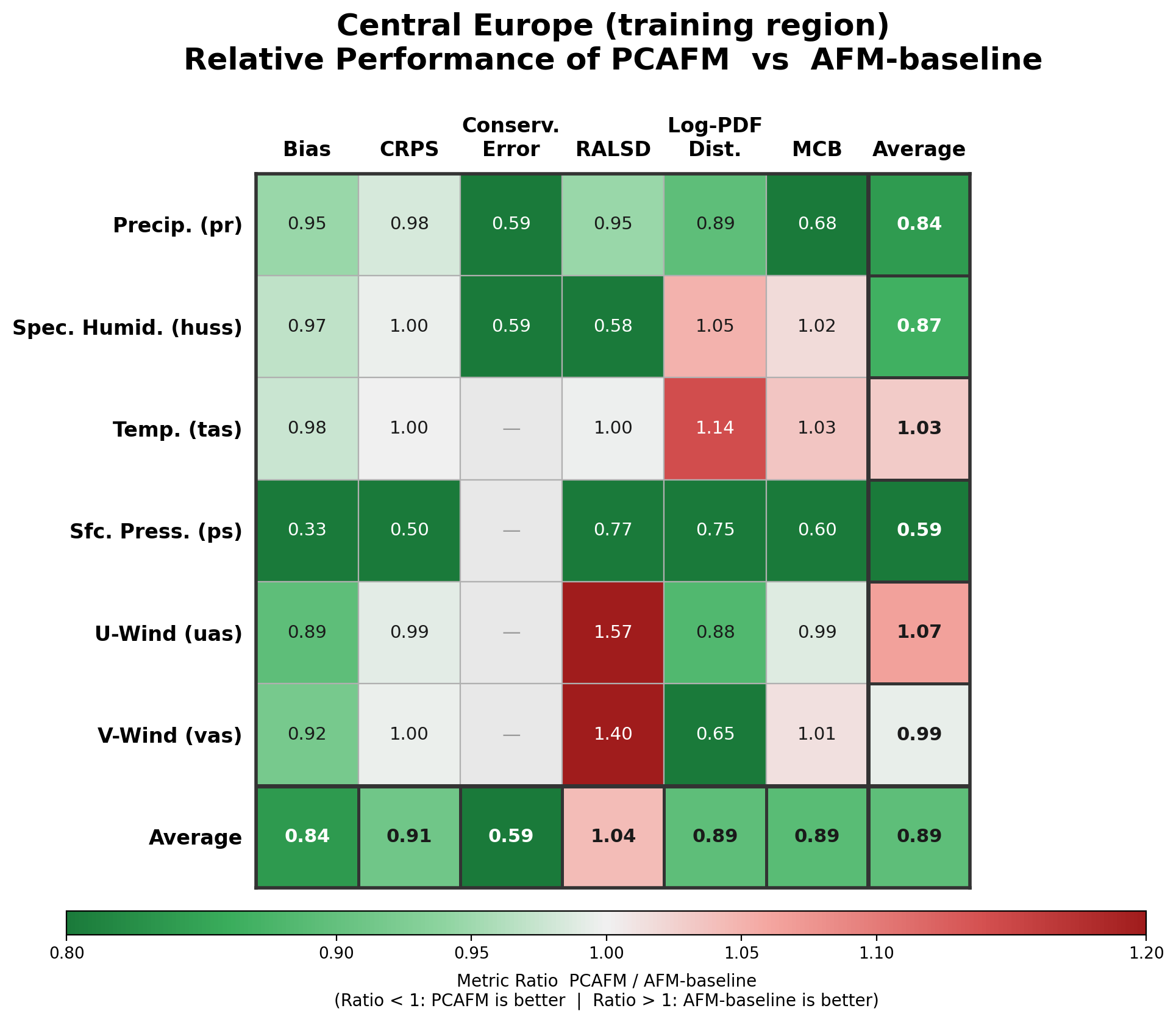}
    \caption{Relative performance of PC-AFM versus AFM-baseline for the Central Europe training region. Each cell shows the ratio of PC-AFM to AFM-baseline; values below~1 (green) indicate improvement. Conservation error is not applicable (``--'') for variables without an explicit conservation constraint. Bold entries summarize row and column averages.}
    \label{fig:rel_perf_CE}
\end{figure}

Figure~\ref{fig:rel_perf_CE} summarizes relative performance in the training domain. PC-AFM achieves an aggregate mean ratio of 0.89, an 11\% improvement compared to AFM. The largest gains are for surface pressure (\texttt{ps}, 0.59), precipitation (\texttt{pr}, 0.84), and specific humidity (\texttt{huss}, 0.87). The \texttt{ps} improvement is notable as no conservation constraint is applied to that variable, pointing to cross-variable regularization of the joint output distribution through the shared denoiser.

\begin{figure}
    \centering
    \includegraphics[width=\textwidth]{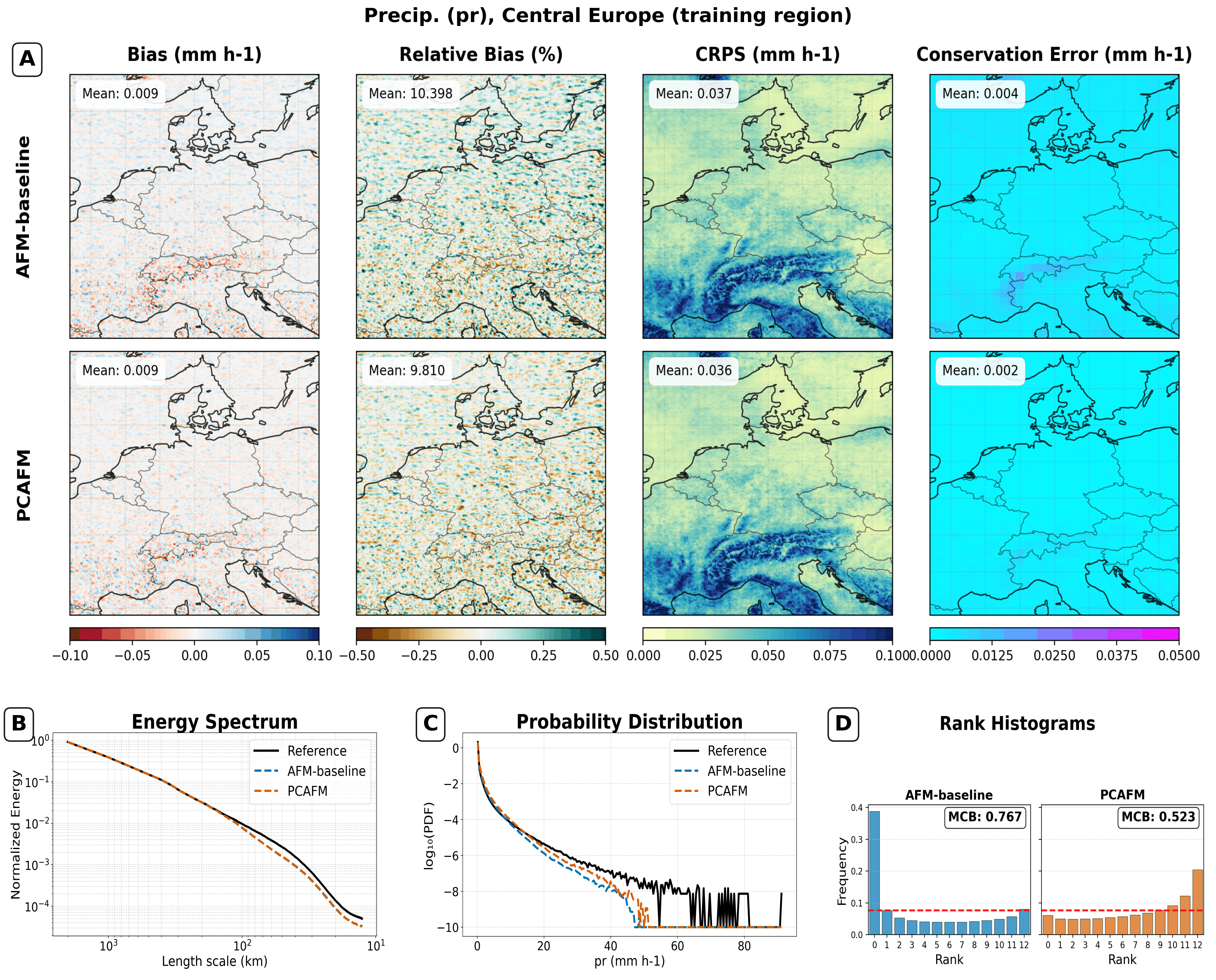}
    \caption{Precipitation (\texttt{pr}) evaluation for the Central Europe training region. (A) Spatial maps of time-mean bias, relative bias, CRPS, and conservation error for AFM-baseline (top) and PC-AFM (bottom). (B) Radially averaged power spectral density. (C) Log-transformed marginal PDF. (D) Rank histograms with MCB. PC-AFM halves the conservation error and improves ensemble calibration (MCB: 0.767 to 0.523) while maintaining comparable bias and CRPS.}
    \label{fig:pr_panel_CE}
\end{figure}

For the directly constrained variables, conservation errors are strongly reduced for precipitation and specific humidity (by ~41 \% for both variables), while CRPS and bias remain essentially unchanged (maps shown in Figures~\ref{fig:pr_panel_CE} and S1. Figure~\ref{fig:pr_panel_CE} shows that ensemble calibration improves notably: precipitation MCB decreases from 0.767 to 0.523. Both models reproduce the reference energy spectrum closely. Wind components and temperature show near-neutral overall performance; wind RALSD is modestly degraded in this domain (1.57 and 1.40 for \texttt{uas} and \texttt{vas}) without an accompanying CRPS degradation, as discussed in Section~\ref{sec:summary}.

\begin{figure}
    \centering
    \includegraphics[width=0.55\textwidth]{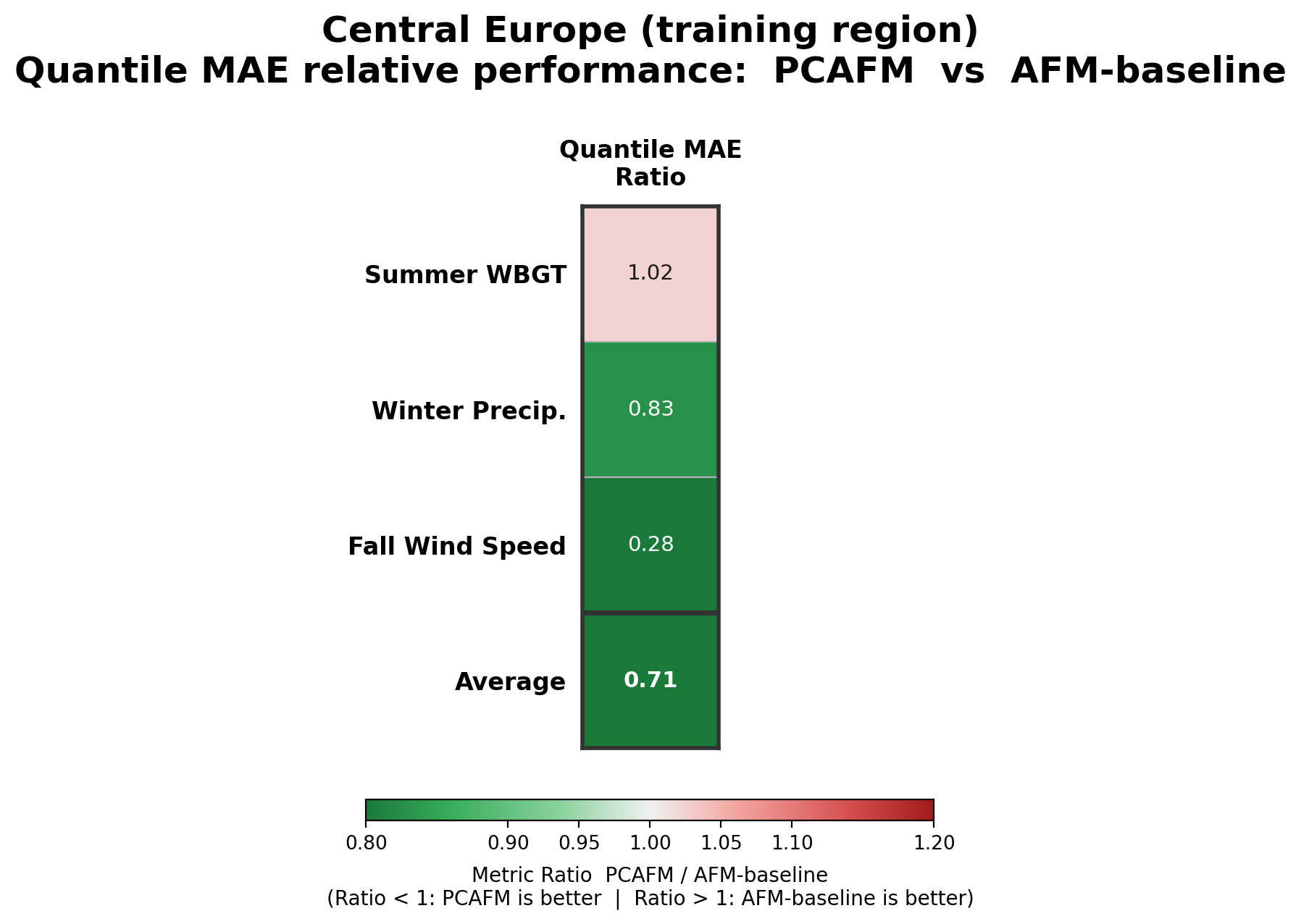}
    \caption{Quantile MAE relative performance (PC-AFM / AFM-baseline) for impact-relevant diagnostics in the Central Europe training region. Average ratio: 0.71.}
    \label{fig:quantile_CE}
\end{figure}

Quantile MAE (Figure~\ref{fig:quantile_CE}) shows a 29\% aggregate improvement (average ratio 0.71). The largest gain is for fall wind speed extremes (ratio 0.28). This is a notable result given that wind is not directly constrained. This could point to cross-variable regularization through the jointly predicted denoiser output. Winter precipitation improves more moderately (0.83). Summer WBGT is near-neutral (1.02). As a compound index derived from \texttt{tas},
\texttt{huss}, and \texttt{ps}, it is possible that modest distributional improvements in the individual variables partially cancel in the derived index.

Supplementary panels for all remaining variables are in Figures~S1, S4, S7, S10, and S13.

\subsection{Geographic Generalization: Northern Europe Region}
\label{sec:results_northerneurope}

\begin{figure}
    \centering
    \includegraphics[width=0.85\textwidth]{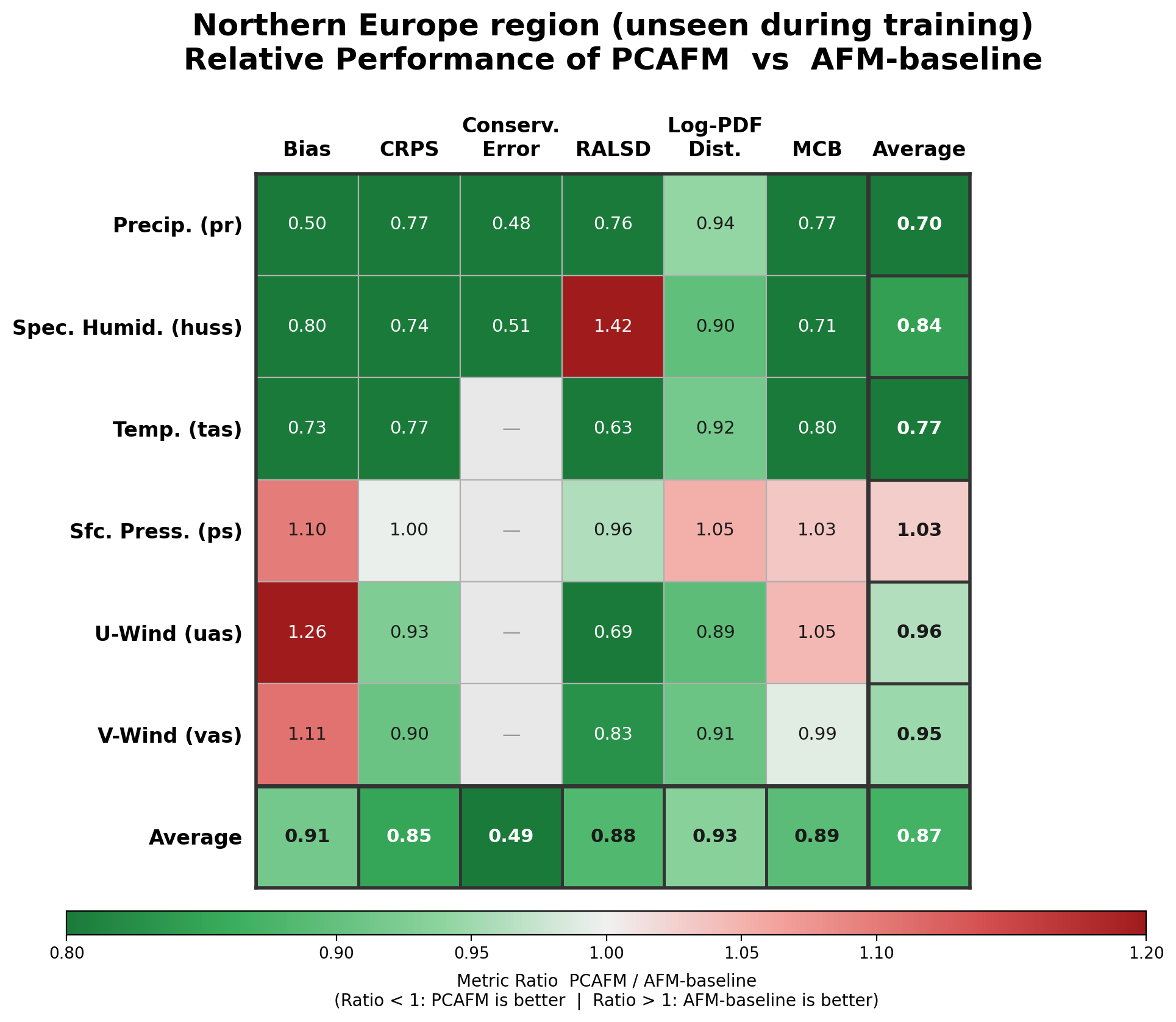}
    \caption{Relative performance of PC-AFM versus AFM-baseline for the Northern Europe region (unseen during training). Layout as in Figure~\ref{fig:rel_perf_CE}.}
    \label{fig:rel_perf_SCA}
\end{figure}

\begin{figure}
    \centering
    \includegraphics[width=\textwidth]{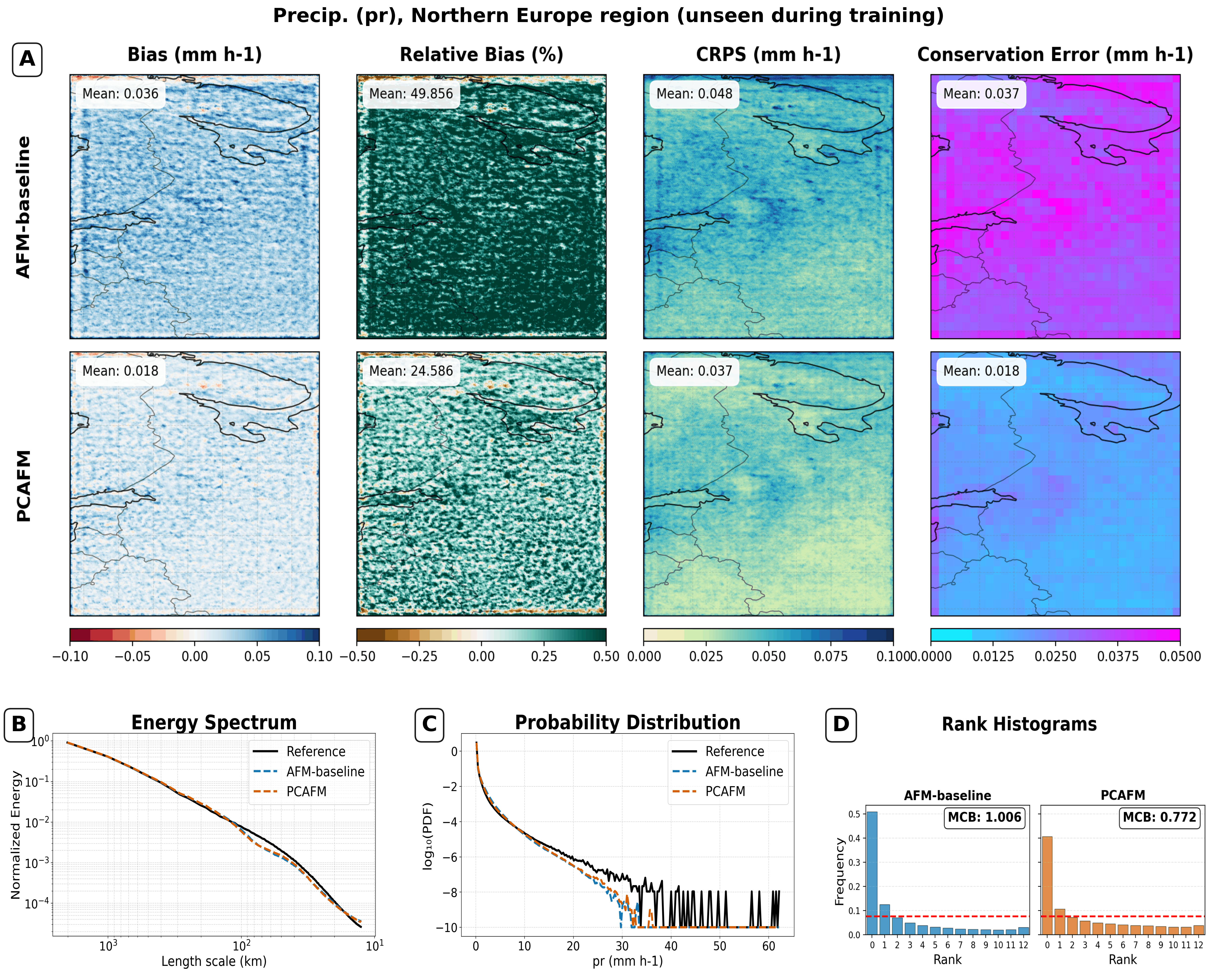}
    \caption{Precipitation evaluation for the Northern Europe region (unseen during training). Layout as in Figure~\ref{fig:pr_panel_CE}. PC-AFM halves the wet bias (49.9\% to 24.6\%), reduces CRPS by 23\%, and cuts conservation error from 0.037 to 0.018~mm~h$^{-1}$. Ensemble calibration improves substantially (MCB: 1.006 to 0.772).}
    \label{fig:pr_panel_SCA}
\end{figure}

\begin{figure}
    \centering
    \includegraphics[width=0.55\textwidth]{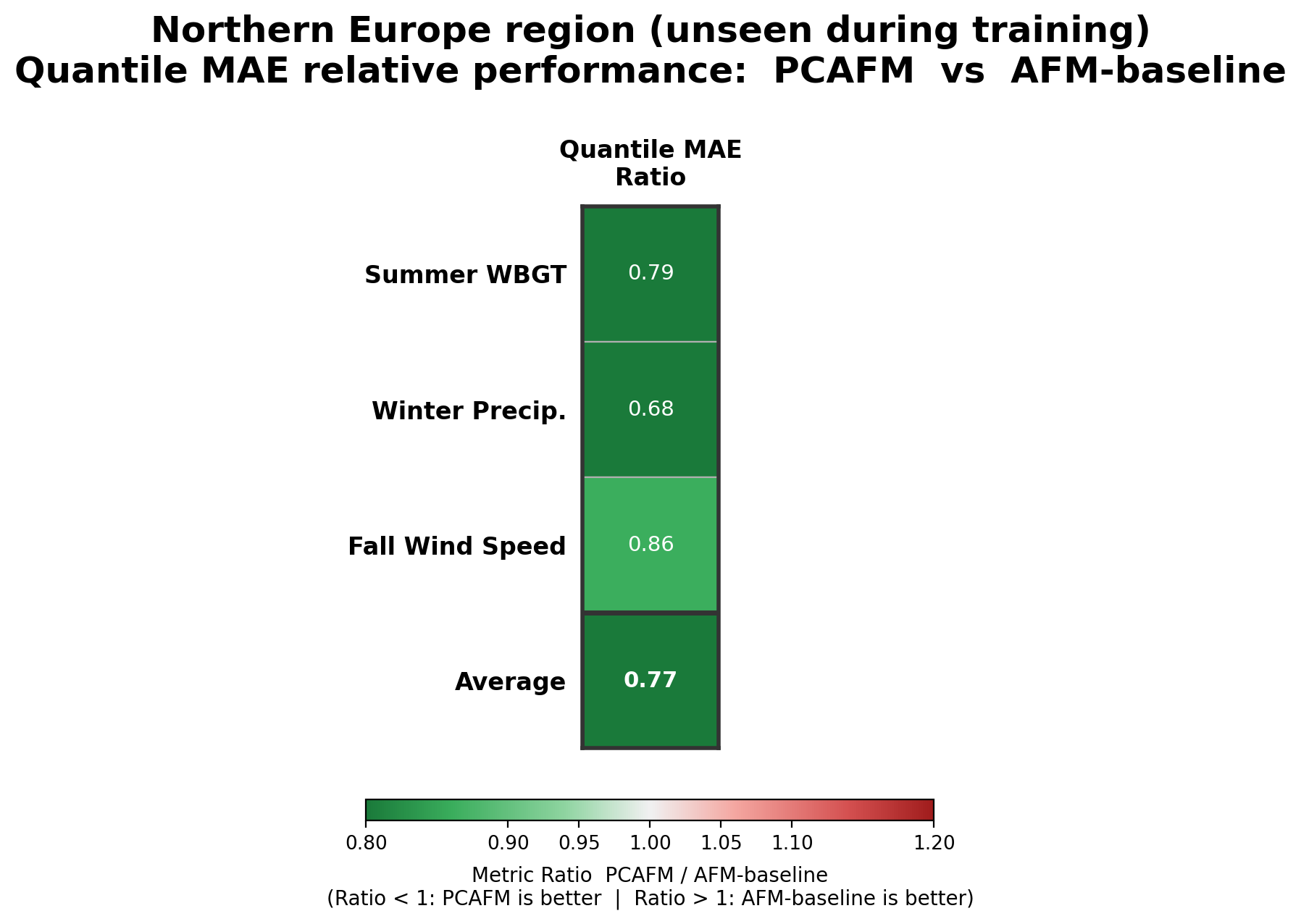}
    \caption{Quantile MAE relative performance for the Northern Europe region (unseen during training). Average ratio: 0.77.}
    \label{fig:quantile_SCA}
\end{figure}

The Northern Europe domain represents the largest distributional shift from the training region, and it is here that the conservation constraint provides the most compelling evidence of its utility. PC-AFM achieves an aggregate ratio of 0.87 (Figure~\ref{fig:rel_perf_SCA}), and conservation gains amplify relative to the training domain: 52\% for precipitation and 49\% for humidity. In Figure~\ref{fig:pr_panel_SCA}, AFM-baseline exhibits a systematic wet bias of 49.9\% in relative precipitation over Northern Europe, driven by overapplying Central European precipitation statistics to a colder climate regime. The conservation constraint halves this bias (49.9\% to 24.6\%) and reduces CRPS by 23\%, providing a physically grounded correction without any information about the target climate at inference time. Temperature also improves strongly (ratio 0.77; Figure~S9). Quantile MAE improves by 23\% on average, including 32\% for winter precipitation extremes (Figure~\ref{fig:quantile_SCA}). Wind bias is modestly degraded (\texttt{uas} ratio 1.26), a cross-variable interaction discussed in Section~\ref{sec:summary}.

Supplementary panels for all remaining variables are in Figures~S3, S6, S9, S12, and S15.

\subsection{Geographic Generalization: Iberian Peninsula}
\label{sec:results_iberia}

\begin{figure}
    \centering
    \includegraphics[width=0.85\textwidth]{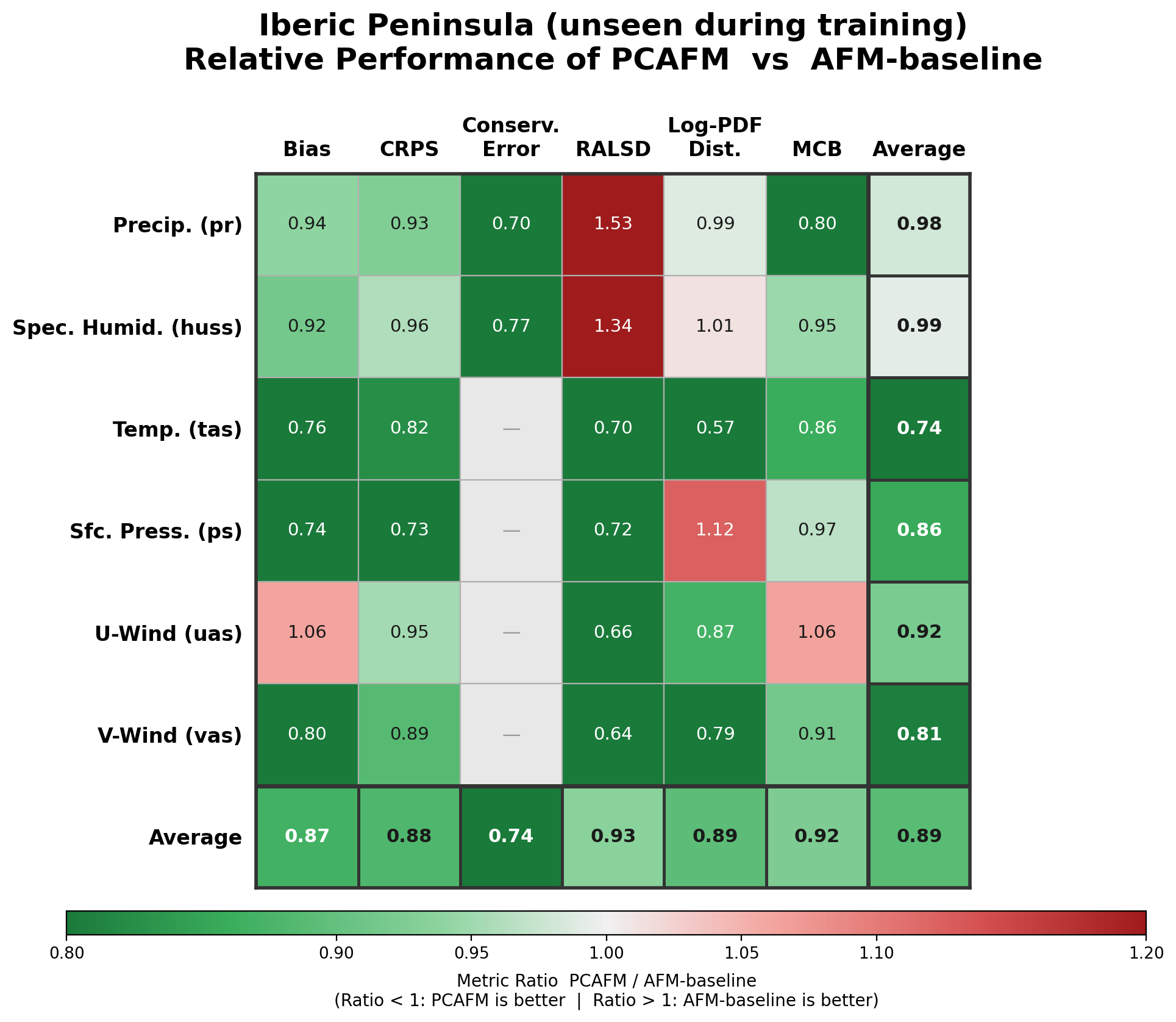}
    \caption{Relative performance of PC-AFM versus AFM-baseline for the Iberian Peninsula (unseen during training). Layout as in Figure~\ref{fig:rel_perf_CE}.}
    \label{fig:rel_perf_IBE}
\end{figure}

\begin{figure}
    \centering
    \includegraphics[width=\textwidth]{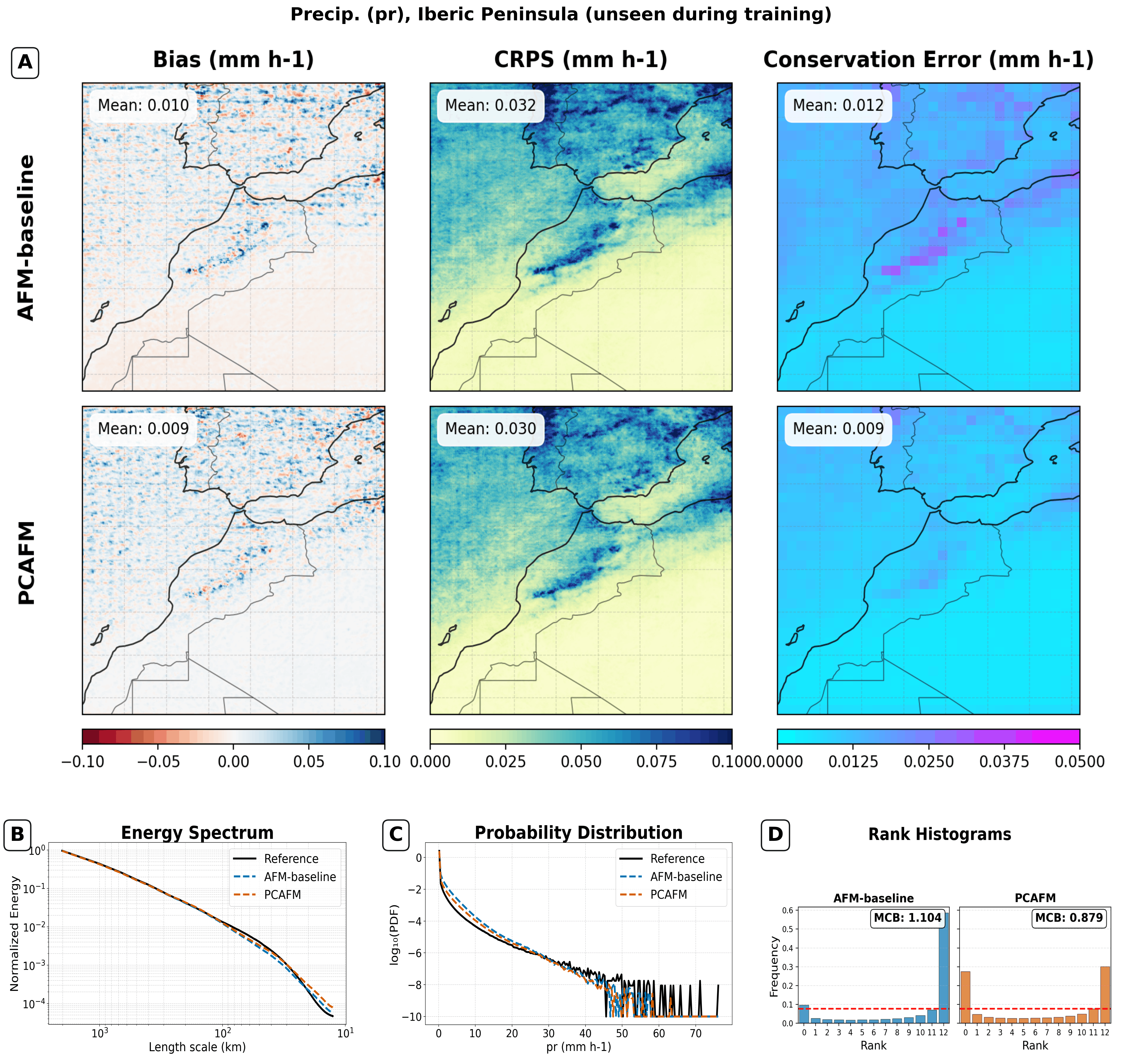}
    \caption{Precipitation evaluation for the Iberian Peninsula (unseen during training). Layout as in Figure~\ref{fig:pr_panel_CE}. PC-AFM reduces conservation error (0.012 to 0.009~mm~h$^{-1}$) and improves ensemble calibration (MCB: 1.104 to 0.879).}
    \label{fig:pr_panel_IBE}
\end{figure}

\begin{figure}
    \centering
    \includegraphics[width=0.55\textwidth]{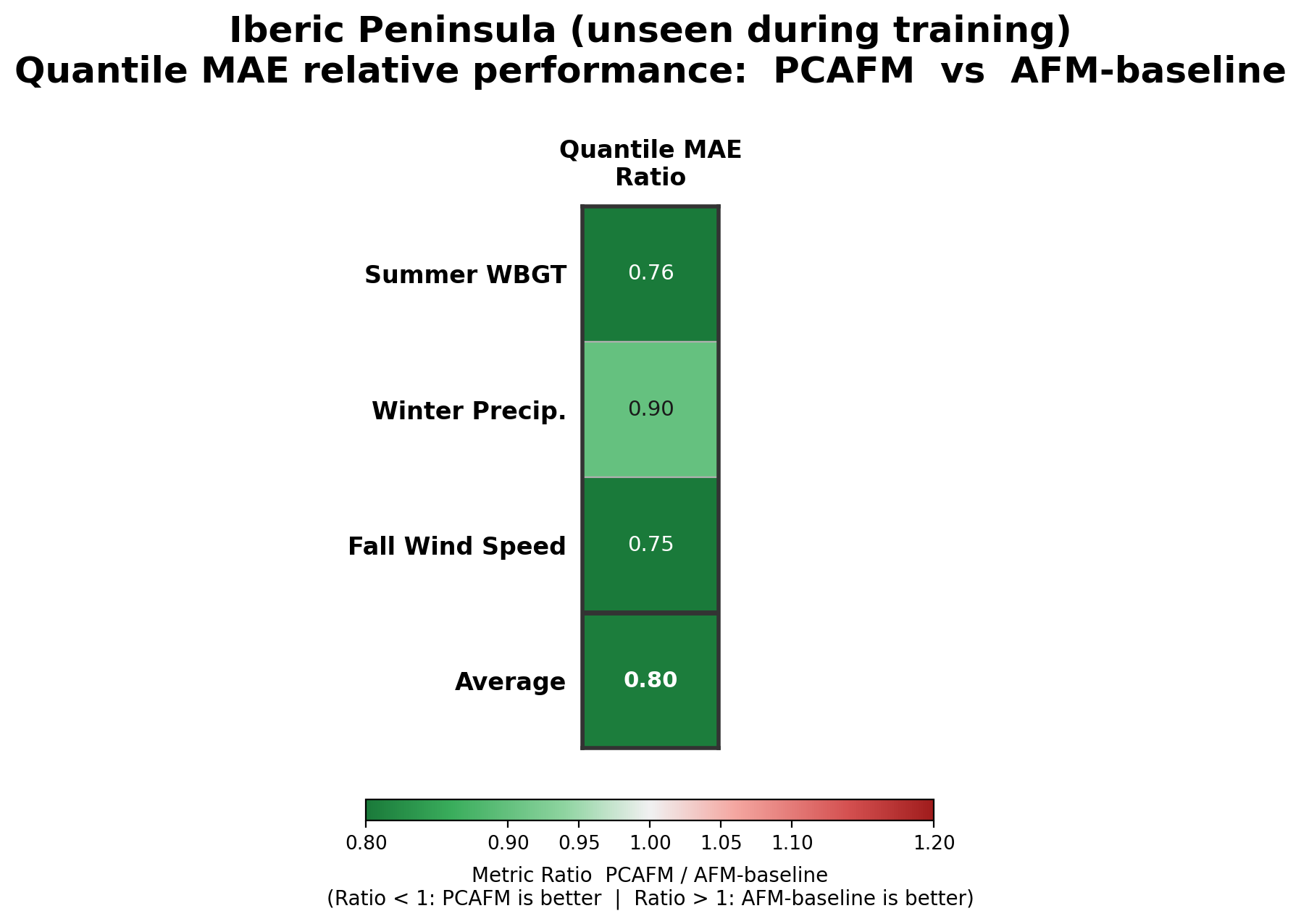}
    \caption{Quantile MAE relative performance for the Iberian Peninsula (unseen during training). Average ratio: 0.80.}
    \label{fig:quantile_IBE}
\end{figure}

PC-AFM achieves an aggregate ratio of 0.89 on the Iberian Peninsula (Figure~\ref{fig:rel_perf_IBE}), matching in-domain performance despite this region being withheld from training. The largest gains are for temperature (\texttt{tas}, 0.74) and surface pressure (\texttt{ps}, 0.86), and conservation errors for the constrained variables are reduced by 30\% and 23\%. Precipitation CRPS improves modestly (0.032 to 0.030~mm~h$^{-1}$) and ensemble calibration improves substantially (MCB: 1.104 to 0.879; Figure~\ref{fig:pr_panel_IBE}). Spectral fidelity of moisture fields degrades modestly in this out-of-distribution regime (RALSD 1.53 for precipitation), reflecting a conservation-spectral trade-off discussed in Section~\ref{sec:summary}. Quantile MAE improves by 20\% on average (Figure~\ref{fig:quantile_IBE}).

Supplementary panels for all remaining variables are in Figures~S2, S5, S8, S11, and S14.

\subsection{Ablation Study}
\label{sec:results_ablation}

We compare four training variants against a matched AFM-baseline at 2~million samples (AFM-baseline-2M) to isolate the contribution of each PC-AFM component: PCAFM-Standard (conservation losses, no gradient surgery, with $\lambda_{\text{phys}} =1$), PCAFM-ConFIG (adds ConFIG gradient surgery), PCAFM-ConFIG-DW (adds noise-level downweighting), and PCAFM-2M (full configuration at 2 millions steps). Results are summarized in Figures~S16--S21.

Without gradient surgery, PCAFM-Standard causes a net regression in Central Europe
(aggregate ratio 1.08), driven by a factor-of-two degradation in surface pressure
(\texttt{ps}, ratio 2.01). This is a direct signature of gradient conflict: the shared denoiser receives opposing updates from the conservation and reconstruction objectives, and the unconstrained variable \texttt{ps} absorbs the inconsistency. ConFIG eliminates this failure immediately, restoring the aggregate to 0.82 and the \texttt{ps} ratio to 0.77. Noise-level downweighting provides additional targeted gains for the constrained variables in-domain (\texttt{huss} ratio 0.65 vs.\ 0.82 for ConFIG alone) but offers limited further benefit out-of-distribution. PCAFM-2M achieves the most consistent
generalization across both withheld regions (aggregate 0.96 on Iberia, 0.89 on Northern Europe) and over quantile MAE. The \texttt{ps} degradation visible at 2 million steps resolves with continued training to 6 million steps (average ratio reduced from 1.51 to 0.59; Section~\ref{sec:results_overall}).

\section{Summary}
\label{sec:summary}

We presented PC-AFM, a generative framework for climate downscaling that extends the adaptive flow matching architecture of \citeA{fotiadis_stochastic_2024} with physics-constrained conservation losses and ConFIG gradient surgery. More broadly, PC-AFM contributes to the emerging paradigm of AI-empowered multiscale climate modeling \cite{eyring_ai-empowered_2024}, in which physically consistent generative models bridge the gap between global simulations and the kilometer-scale information required for climate change mitigation and adaptation.

The most consistent benefit of PC-AFM is the reduction of cross-scale conservation error, and this benefit amplifies under distributional shift. In the training domain, conservation errors for precipitation and humidity are reduced by approximately 40\%, ensemble calibration improves substantially, and aggregate performance improves by around 10\% over the unconstrained AFM baseline. Surface pressure, though not directly constrained, shows strong overall improvement (ratio 0.59), reflecting beneficial cross-variable regularization. As the denoiser predicts all six channels jointly, conservation corrections to precipitation and specific humidity fields propagate through the learned multivariate correlations to pressure and temperature. Conservation constraints substantially improve out-of-distribution generalization. In the Northern Europe domain, PC-AFM halves a large systematic precipitation wet bias (49.9\% to 24.6\%) and reduces CRPS by 23\%, with conservation error reductions amplifying to 52\%. Extreme-quantile accuracy improves by 20--23\% across both withheld regions. These results are consistent with the theoretical analysis of \citeA{baldan_physics_2026}, who found that flow-matching models with physical constraints define a Pareto frontier between physical fidelity and distributional accuracy. In the in-distribution setting, the constraint primarily reduces conservation error while leaving CRPS unchanged, because the model has already learned approximate conservation implicitly. Under distribution shift, the constraint becomes the dominant source of generalization, correcting systematic biases that the model would otherwise inherit from training-domain statistics.

ConFIG gradient surgery is essential to realize these gains without affecting 
reconstruction quality. Without gradient surgery, the competing objectives produce conflicting updates leading to suboptimal compromises regardless of the chosen loss weight. The noise-level-dependent weighting concentrates the conservation penalty where it is physically meaningful: at low noise levels, where the denoiser output approximates an actual physical field. The warmup period serves a complementary role, letting the model establish a reasonable mapping before the conservation constraints play a role in the learned representations.

A notable cost is modest spectral degradation of moisture fields in out-of-distribution regions. The conservation penalty suppresses spatial variance at intermediate scales when the model's fine-scale structure is poorly suited to the target climate, consistent with findings by \citeA{saccardi_assessing_2025}. The wind bias degradation in Northern Europe (\texttt{uas} ratio 1.26) may reflect an adverse cross-variable interaction, whereby conservation corrections applied to precipitation and humidity propagate through the learned multivariate correlations and introduce a slight offset in the correlated wind fields.

Several directions remain for future work. Spectral bias persists across all tested regions; future work could explore spectral-aware loss terms or wavelet-based constraints to address this. Conservation is currently enforced only at the surface level; incorporating multi-level or thermodynamic constraints could improve variable consistency. Time steps are downscaled independently, so temporal conditioning or autoregressive generation would be needed for applications requiring coherent sequences. While we demonstrate generalization across geographically distinct climate regimes, the evaluation does not address temporal out-of-distribution shift; testing PC-AFM on future scenario simulations would be required to validate robustness under transient climate change. The framework could also be extended to additional conserved quantities such as energy balance or surface wind divergence near complex topography.

A more fundamental limitation concerns the validity of the conservation assumption itself. The PC-AFM conservation constraint (Equation~\ref{eq:conservation}) rests on the premise that the area-weighted mean of the true high-resolution field exactly equals the coarse-scale input, which holds by construction in our experimental setup. In operational downscaling, however, the coarse-scale driving fields come from an independent GCM whose internal dynamics evolve separately, and applying the conservation constraint could force the downscaled field toward a biased large-scale mean. Extending PC-AFM to real-world GCM-driven downscaling would therefore require either a relaxed soft constraint with uncertainty-aware weighting, or an explicit correction step that accounts for the systematic offset between the driving GCM and the reference high-resolution climatology.


\section*{Open Research Section}

The nextGEMS cycle~3 simulation data are available from \url{https://doi.org/10.5281/zenodo.15024577}.
The preprocessed training dataset, code for training and evaluating PC-AFM (including the ESMValTool diagnostic recipes used in this study), and trained model weights for all evaluated variants will be archived and made publicly available once the paper is accepted.

\section*{Conflict of Interest Declaration}

The authors declare there are no conflicts of interest for this manuscript.

\acknowledgments
KD, AP, VE and PG have received funding for this study from the European Research Council (ERC) Synergy Grant “Understanding and Modelling the Earth System with Machine Learning (USMILE)” within the EU Horizon 2020 research and innovation program (grant agreement no. 855187). VE was additionally supported by the Deutsche Forschungsgemeinschaft (DFG, German Research Foundation) through the Gottfried Wilhelm Leibniz Prize (reference no. EY 22/2-1). PG acknowledges National Science Foundation Learning the Earth with AI and Physics (LEAP) grant Award \#2019625-STC. We acknowledge the computational resources of the Deutsches Klimarechenzentrum (DKRZ, Hamburg, Germany) used in this study (grant no. 1179).

\bibliography{refs}

\end{document}